%% file: fastsscflip.tex
\documentclass[conference]{IEEEtran}
\usepackage{etex}
\usepackage{ifpdf}
\ifpdf
\usepackage[draft,pdfborder={0 0 0}]{hyperref}
\fi

\usepackage{cite}
\usepackage[demo]{graphicx}

\usepackage[cmex10]{amsmath}
\usepackage{nicefrac}

\newcommand{\mvec}[1]{\bm{#1}}

\usepackage{bm}
\usepackage[varg]{txfonts}
\let\mathbb=\varmathbb
\DeclareSymbolFont{letters}{OML}{ztmcm}{m}{it}

\usepackage{array}
\usepackage[caption=false,font=footnotesize]{subfig}

\usepackage{tabularx}
\usepackage{multirow}
\usepackage{multicol}
\usepackage{dcolumn}
\usepackage{booktabs}

\usepackage{stfloats}

\usepackage{tikz,pgf}
\usepackage{pgfplots}
\pgfplotsset{compat=1.14}
\usetikzlibrary{shapes,positioning,arrows,decorations.markings,fit,calc,patterns,spy,plotmarks}

\usepackage{glossaries}

\usepackage{balance}

\newcommand{\sgn}[1]{\text{sgn}(#1)}

\title{Fast-SSC-Flip Decoding of Polar Codes}

\author{\IEEEauthorblockN{
Pascal Giard and Andreas Burg}\vspace{2pt}
  \IEEEauthorblockA{Telecommunications Circuits Laboratory\\Ecole polytechnique f\'ed\'erale de Lausanne (EPFL), 1015 Lausanne VD, Switzerland\\Email: \{pascal.giard,andreas.burg\}@epfl.ch}}

\begin{document}

\newacronym[plural=CRCs]{crc}{CRC}{cyclic redundancy check}
\newacronym{fer}{FER}{frame-error rate}
\newacronym{ber}{BER}{bit-error rate}
\newacronym{ldpc}{LDPC}{low-density parity-check}
\newacronym{sc}{SC}{successive-cancellation}
\newacronym{bp}{BP}{belief-propagation}
\newacronym{bpsk}{BPSK}{binary phase-shift keying}
\newacronym{awgn}{AWGN}{additive white Gaussian noise}
\newacronym[plural=LLRs,firstplural=log-likelihood ratios (LLRs)]{llr}{LLR}{log-likelihood-ratio}
\newacronym{ssc}{SSC}{simplified \gls{sc}}
\newacronym[first=fast-SSC]{fastssc}{fast-SSC}{fast-\gls{ssc}}
\newacronym{scl}{SCL}{successive-cancellation list}
\newacronym{scf}{SCF}{successive-cancellation flip}
\newacronym{spc}{SPC}{single-parity-check}
\newacronym[plural=CCs,firstplural=clock cycles (CCs)]{cc}{CC}{clock cycle}
\newacronym{wc}{W.-C.}{worst-case}

\maketitle

\begin{abstract}
  Polar codes are widely considered as one of the most exciting recent discoveries in channel coding. For short to moderate block lengths, their error-correction performance under list decoding can outperform that of other modern error-correcting codes.
  However, high-speed list-based decoders with moderate complexity are challenging to implement. 
  Successive-cancellation (SC)-flip decoding was shown to be capable of a competitive error-correction performance compared to that of list decoding with a small list size, at a fraction of the complexity, but suffers from a variable execution time and a higher worst-case latency.
  In this work, we show how to modify the state-of-the-art high-speed SC decoding algorithm to incorporate the SC-flip ideas. The algorithmic improvements are presented as well as average execution-time results tailored to a hardware implementation. The results show that the proposed fast-SSC-flip algorithm has a decoding speed close to an order of magnitude better than the previous works while retaining a comparable error-correction performance.
\end{abstract}

\section{Introduction}
\label{sec:intro}
Polar codes made it into 3GPP's next-generation mobile-communication standard (5G) due to their excellent error-correction performance under \gls{scl} decoding \cite{3GPPRANPolar}. However, implementing high-throughput \gls{scl} decoders while retaining a moderate complexity is challenging since the decoder speculatively explores multiple candidate solutions in parallel of which a majority is, in the end, abandoned. As an alternative, Afisiadis \textit{et al.} proposed the low-complexity \gls{scf} decoding algorithm \cite{Afisiadis2014} which explores candidate solutions sequentially, therefore avoiding many unnecessary computations. They showed that, for a 1024-bit polar code with a code rate of $\nicefrac{1}{2}$ and a 16-bit \gls{crc}, \gls{scf} could match the error-correction performance of \gls{scl} decoding with a small list size ($L=2$) if a sufficient number of trials were attempted ($T=32$). However, being sequential in nature, the instantaneous decoding throughput of \gls{scf} decoding is variable and the average throughput depends on the channel signal-to-noise ratio. Nevertheless, it was shown in \cite{Giard_JETCAS_2017} that, under reasonable conditions, it had much better average throughput and a significantly lower average decoding complexity in terms of the area-time product than \gls{scl} decoding.

\subsubsection*{Contributions}
In this paper, we show how to merge the \gls{scf} decoding algorithm with the state-of-the-art high-speed \gls{sc}-based decoding algorithm that decomposes polar codes into constituent codes. We introduce decision-\gls{llr} calculations, required for \gls{scf} decoding, tailored to the various constituent-code types. We show that the new decoding algorithm has an error-correction performance that is either virtually the same or very close to that of the original \gls{scf} algorithm. Hardware implementation considerations are discussed, and the execution time of the proposed algorithm is compared with that of the only \gls{scf}-decoder implementation from the literature.

\subsubsection*{Outline}
The remainder of this paper starts with Section~\ref{sec:bg} which provides background about polar-code construction and encoding, polar-code representations and decomposition in constituent codes, and the decoding algorithms on which this work is built.
Section~\ref{sec:algo} describes the proposed new algorithm and the decision-\gls{llr} equations for the various constituent-code types. Hardware considerations are discussed in Section~\ref{sec:hw}, where an average execution-time comparison against the state of the art is also presented. Finally, Section~\ref{sec:conclusion} concludes this paper.

\section{Background}
\label{sec:bg}
\subsection{Construction and Encoding}
Polar codes are linear block codes, i.e., the encoding process implies the linear transformation of a vector of bits. This transformation is structured in a way that results in a polarization effect, as its length tends to infinity, where some of the encoded bits can be decoded perfectly while the others become completely unreliable.

In particular, in matrix form, a polar code of length $N$ can be obtained as
\begin{align}
  \label{eq:encoding}
	\mvec{x} = & \mvec{u}\mvec{F}^{\otimes n},	\qquad \mvec{F} = \begin{bmatrix} 1 & 0 \\ 1 & 1 \end{bmatrix},
\end{align}
where $n \triangleq \log _2 N$, $\mvec{u}$ is the vector of bits to be encoded, and $\mvec{F}^{\otimes n}$ is the $n^{\text{th}}$ Kronecker product of $\mvec{F}$ and $\mvec{F}^{\otimes 1}=\mvec{F}$. To obtain an $(N,\,k)$ polar code of rate $R=\nicefrac{k}{N}$, the $k$ most-reliable bit locations in $\mvec{u}$ hold the information bits while the other $N-k$ bits, called frozen bits, are set to a predetermined value (usually 0). The bit-location reliabilities depend on the channel type and condition. Many methods have been proposed to calculate these reliabilities; we use that of Tal and Vardy \cite{Tal2011a}.

\subsection{Representations and Constituent Codes}

\begin{figure}[t]
  \begin{minipage}{0.55\columnwidth}
    \centering
    \subfloat[Graph]{\label{fig:pc8}\hspace{-15pt}\resizebox{1.15\columnwidth}{!}{\input{figures/pc8.tex}}}
  \end{minipage}%
  \begin{minipage}{0.45\columnwidth}
    \centering
    \subfloat[SC Decoder Tree]{\label{fig:sc-tree}\rotatebox{90}{\input{figures/sc-tree.tex}}}
    \vspace{-2pt}
    \subfloat[Fast-SSC Decoder Tree]{\makebox[\columnwidth][c]{\label{fig:fastssc-tree}\rotatebox{90}{\input{figures/fastssc-tree.tex}}}}
  \end{minipage}
  \caption{Graph and decoder-tree representations of an $(8,\,5)$ polar code.}
\end{figure}
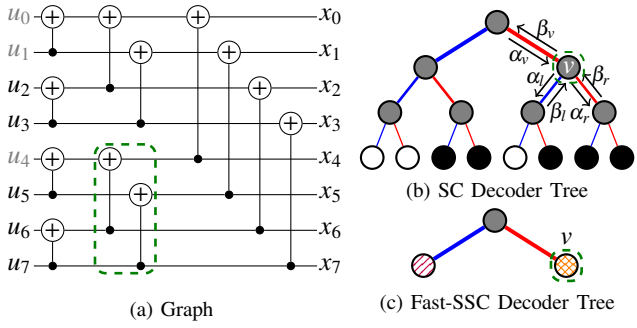

In addition to the matrix form, polar codes can be represented as a graph. Fig.~\ref{fig:pc8} shows such a representation for an $(8, 5)$ polar code, where \tikz{\node[draw,circle,minimum size=0.1cm,inner sep=0pt] at (0,0) {\small $+$};} are modulo-2 additions, the grayed $u_i$'s with $i \in \{0,1,4\}$ hold frozen bits, and the black $u_i$'s with $i \in \{2, 3, 5, 6, 7\}$ hold the information bits. Encoding is done by propagating the vector $\mvec{u}$, in that graph, from left to right.

Also from Fig.~\ref{fig:pc8}, notice how polar codes are built recursively: the first half ($x_0^3 \triangleq \left[x_0, x_1, x_2, x_3\right]$) of a polar code of length $N=8$ is the result of the element-wise modulo-2 addition of two polar codes of length $N_v=\nicefrac{N}{2}=4$, and the other half ($x_4^7$) corresponds to the second polar code of length $N_v=4$. In other words, any polar code of length $N$ can be seen as a composition of two smaller constituent (polar) codes of length $N_v=\nicefrac{N}{2}$ with $k_v$ information bits. Taking into consideration the frozen-bit locations, many of these constituent codes can be considered as block codes with a special structure rather than as polar codes. As will be briefly reminded in Section~\ref{sec:fastssc} below, this structure can then be exploited to use dedicated decoding algorithms which are more efficient than the generic decoding algorithm for polar codes.

Alternatively to the graph representation, it was shown in \cite{Alamdar-Yazdi2011} that polar codes can also be represented as binary trees, or decoder trees, where the white and black leaf nodes correspond to frozen-bit and information-bit locations, respectively. Fig.\ref{fig:sc-tree} shows the decoder tree representation of the same $(8,\,5)$ polar code illustrated as a graph in Fig.~\ref{fig:pc8}. The part of the graph circled by a green dashed line in Fig.~\ref{fig:pc8} corresponds to the node $v$ of width $N_v$ circled in the same way in Fig.~\ref{fig:sc-tree}.

\subsection{Successive-Cancellation and Fast-SSC Decoding}\label{sec:fastssc}
The \gls{sc} decoding algorithm as initially proposed~\cite{Arikan2009} proceeds by visiting the decoder-tree representation---e.g., Fig.~\ref{fig:sc-tree}---sequentially, from top to bottom, from left to right, successively estimating $\bm{\hat{u}}$ at the leaf nodes, from the noisy channel values. Visiting a left edge (blue) on this representation, the \gls{sc} algorithm can calculate the soft-input \glspl{llr} to the child node $\alpha_l$ with the min-sum approximation \cite{Leroux2011}
\begin{equation}\label{eqn:sc:f}
\alpha_l[i] = \sgn{\alpha_v[i]\alpha_v[i + \nicefrac{N_v}{2}]} \min(|\alpha_v[i]|, |\alpha_v[i + \nicefrac{N_v}{2}]|),
\end{equation}
where $\alpha_v$ and $N_v$ are respectively the \glspl{llr} and node length from the parent node. At the root node, the channel \glspl{llr} are used. Once a leaf node is reached, a bit $\hat{u}_i$ (for a non-systematic polar code) is estimated as
\begin{equation}\label{eqn:sc:estu}
\hat{u}_i = \begin{cases}
0\text{,} & \text{when } \alpha_v \geq 0~\text{or}~i \in \mathcal{F};\\
1\text{,} & \text{otherwise,}
\end{cases}
\end{equation}
where $\mathcal{F}$ is the set of frozen-bit indices.\footnote{In the case of a systematic polar code under \gls{sc} decoding, the estimated-bit vector is obtained at the end of the decoding process by calculating $\mvec{\hat{u}}_0^{N-1}\mvec{F}^{\otimes n}$.}

Visiting a right edge (red), the \glspl{llr} to the child node $\alpha_r$ can be calculated \cite{Leroux2011} as
\begin{equation}\label{eqn:sc:g}
\alpha_r[i] = \begin{cases}
\alpha_v[i + \nicefrac{N_v}{2}] + \alpha_v[i]\text{,} & \text{when } \beta_l[i] = 0;\\
\alpha_v[i + \nicefrac{N_v}{2}] - \alpha_v[i]\text{,} & \text{otherwise},
\end{cases}
\end{equation}
where $\beta_l$ is the bit-estimate vector generated by the left sibling in the decoder-tree. If the left sibling is a leaf node, its estimated-bit value $\hat{u}_i$ is used as the $\beta_l$. Otherwise, the estimed-bit vector $\beta_v$ at a node $v$ is calculated as
\begin{equation}
\beta_v[i] =
  \begin{cases}
    \beta_l[i]\oplus \beta_r[i], & \text{when}~i < \nicefrac{N_v}{2}\\
    \beta_r[i+\nicefrac{N_v}{2}], & \text{otherwise},
  \end{cases}
  \label{eq4}
\end{equation}
where $\beta_l$ and $\beta_r$ are the bit-estimate vectors from the left- and right-child nodes, respectively, and $\oplus$ denotes a modulo-2 addition.

Alamdar-Yazdi and Kschischang proposed the \gls{ssc} algorithm where the subtrees solely composed of either frozen (rate-0 codes) or information nodes (rate-1 codes) are not fully traversed \cite{Alamdar-Yazdi2011}. Recognizing more types of constituent codes, specialized algorithms and corresponding dedicated decoders were proposed in \cite{Sarkis_JSAC_2014}. The algorithm described in \cite{Sarkis_JSAC_2014}, and later extended in \cite{Giard_JSPS_2016b} and \cite{Hanif2017}, is referred to as the \gls{fastssc} algorithm. Details on the specialized algorithms of \gls{fastssc} used in this paper are provided alongside the proposed algorithm in Section~\ref{sec:algo} below. \Gls{fastssc} decoders have a throughput that depend on the frozen-bit locations, however it is typically an order of magnitude higher than that of other \gls{sc}-based decoders.

Fig.~\ref{fig:fastssc-tree} shows a decoder tree for the same $(8,\,5)$ polar code as above where the \gls{fastssc} algorithm is applied, i.e., where the tree is pruned by recognizing that the left-hand-side subtree of Fig.~\ref{fig:sc-tree} corresponds to the ML node of \cite{Sarkis_JSAC_2014} (Type-I node in \cite{Hanif2017}) and the right-hand-side subtree to a \gls{spc} node. In Fig.~\ref{fig:fastssc-tree}, the former subtree is replaced by a purple-striped node and the latter with an orange-hatched node.

\subsection{Successive-Cancellation Flip Decoding}
The \gls{scf} decoding algorithm builds upon a slightly-modified \gls{sc} decoding algorithm.
It starts by going through a regular \gls{sc}-decoding pass (first trial). In parallel to the decoding process, a list of (absolute) decision \glspl{llr} associated to each estimated information bit is built. Once \gls{sc} decoding has completed, the embedded \gls{crc} is verified. In case it matches, decoding stops and the estimated codeword is output. Otherwise, another \gls{sc}-decoding pass (second trial) is launched, however, this time once the location of the information bit that corresponds to the least-reliable decision \gls{llr} (from the first trial) is reached, that estimated bit is flipped before resuming \gls{sc} decoding. Once \gls{sc} decoding has completed, the \gls{crc} is verified again. If it matches, the estimated codeword is output otherwise the process is restarted, where the bit corresponding to the second-least-reliable decision \gls{llr} is to be flipped. This process goes on until the maximum number of trials $T_{\max}$ is reached. Note that setting $T_{\max}=1$  corresponds to regular \gls{sc} decoding.

By nature, the latency of \gls{scf} decoding is defined as a multiple of the underlying \gls{sc} decoder
\begin{equation}\label{eqn:latency:scf}
  \mathcal{L}_{\text{SCF}} = T_{\max}\mathcal{L}_{\text{SC}},
\end{equation}
where $\mathcal{L}_{\text{SC}}$ is the latency of the underlying \gls{sc} decoder. Furthermore, both the average and worst-case throughput are functions of the throughput of that same \gls{sc} decoder. Thus, improving the speed of the underlying \gls{sc} decoder also improves the throughput of the \gls{scf} decoder.

\section{Fast-SSC-Flip Decoding}\label{sec:algo}

During the first trial, the \gls{scf} decoding algorithm builds a list that contains the decision \glspl{llr} corresponding to each information bits. That list $\mvec{\lambda}$ is used to determine which information bit to flip in subsequent trials. In the following, we show how the \gls{fastssc} and \gls{scf} algorithms can be merged. 

The \gls{fastssc} algorithm uses dedicated decoders to estimate multiple information bits at a time. These decoders need to be modified to calculate the decision \glspl{llr} required by the \gls{scf} algorithm, and to add support for the bit-flipping procedure. These modifications are described in the following subsections for the essential leaf-node types (constituent codes) that contain information bits. The other leaf-node types that may be encountered can all be expressed as node combinations that include the types covered below.

\subsection{Information Nodes}
Information nodes have length $N_v$ and contain $k_v=N_v$ information bits,  none of their bit locations is frozen. Their \gls{ssc} decoding is a hard decision on the node soft-input \glspl{llr} $\alpha_0^{N_v-1}$. As no parity (frozen) bits are involved, the calculation of the decision \glspl{llr}, $\lambda_d, 0 \leq d < k_v$, remains the same as in the original \gls{scf} algorithm \cite{Afisiadis2014}, i.e., as the absolute value of each $\alpha_i, 0 \leq i < N_v$
\begin{equation}
  \lambda_d=\left|\alpha_d\right| \text{ for } 0 \leq d < k_v, i=d\,.
\end{equation}

In case the decoder has already passed the first trial, and the index of the information bit to be flipped falls within an information node, after decoding the rate-1 code, the bit that corresponds to that index is flipped.

\subsection{Repetition Nodes}
Repetition nodes protect a single information bit ($k_v=1$) by repeating it $N_v$ times at encoding time. Thus, \gls{fastssc} decoding takes a hard decision on the sum of the $N_v$ node input \glspl{llr}. For the decision \gls{llr}, the original \gls{scf} algorithm uses the absolute value of the \gls{llr} that corresponds to the information bit. For a repetition node, both the \gls{sc} and \gls{fastssc} algorithms effectively calculate that \gls{llr} as the sum of all node input \glspl{llr}. Thus, it is proposed that the decision \gls{llr} be calculated in the same way
\begin{equation}
  \label{eq:rep:lambda}
  \lambda_d=\left|\,\sum_{i=0}^{N_v-1}\alpha_i\,\right|,\text{ for }d=k_v-1\,.
\end{equation}

After the first trial, if the index of the information bit to be flipped corresponds to that protected by a repetition code, that bit is flipped after decoding.

\subsection{Birepetition Nodes}
We define as a birepetition node a node where all bit locations are frozen with the exception of the two most-significant positions that carry information bits ($k_v=2$). In the original \gls{fastssc} algorithm \cite{Sarkis_JSAC_2014}, birepetition codes of length $N_v=4$ were decoded by the ML node, whereas longer birepetition codes were decomposed. It was shown in \cite{Hanif2017} that they can be efficiently decoded by recognizing the similarity with repetition nodes, i.e., they can be efficiently decoded as two independent repetition codes. The first repetition code is composed from the even-indexed locations and the second one from the odd-indexed locations. Therefore, the decision-\gls{llr} calculations are
\begin{equation}
  \label{eq:birep:lambda}
  \lambda_d=\left|\,\sum_{i=0}^{\nicefrac{N_v}{2}-1}\alpha_{2i+d}\,\right|, \text{ for } 0 \leq d < k_v\,.
\end{equation}

Similarly to the repetition node, if the index of the information bit to be flipped corresponds to one of the two bits protected by a birepetition code, the corresponding information bit (even or odd indexed) is flipped after decoding.

\subsection{SPC Nodes}
\Gls{spc} nodes are a special type of $(N_v,\,k_v)$ polar codes with $k_v=N_v-1$, where the only frozen bit is in the first location. A maximum-likelihood decoding algorithm for \gls{spc} codes, when a single estimated-bit vector is to be retained, can be summarized as flipping the information bit that corresponds to the least-reliable input \gls{llr} when the parity-check bit is not satisfied \cite{Silverman1954,Sarkis_JSAC_2014}.

The challenge to adapt this type of node to \gls{scf} decoding is to calculate with low complexity meaningful alternative decision \glspl{llr} that take into account the parity constraint. Calculating the exact decision \glspl{llr} as the original \gls{scf} algorithm would involved too many calculations. Thus, we propose an approximation similar to the detection metric update rule for \gls{spc} nodes proposed in \cite{Giard_SIPS_2017}. To this end, we define the decision \glspl{llr} as
\begin{equation}\label{eq:spc:decision}
  \lambda_d=\left|\alpha_i\right| + s(-1)^p\min\left(\left| \alpha_0^{N_v-1} \right| \right), \text{ for } 0 \leq d < k_v, i=d+1,
\end{equation}
where $s$ is a scaling factor, and $p$ is the calculated parity on all $N_v$ input \glspl{llr}, i.e.,

\noindent\begin{minipage}{.55\linewidth}
  \[
    \text{HD}[i] = \begin{cases}
      0, & \text{when } \alpha_i \geq 0;\\
      1, & \text{otherwise,}
    \end{cases}
  \]
\end{minipage}%
\begin{minipage}{.05\linewidth}
  and
\end{minipage}%
\begin{minipage}{.4\linewidth}
  \[
    p=\bigoplus_{i=0}^{N_v-1}\text{HD}[i].
  \]
\end{minipage}\\

The approximations on the decision \glspl{llr} incur an error-correction performance loss. As will be shown in the next section, this loss can be partially compensated for by using the scaling factor $s$ but, more importantly, it becomes negligible as the maximum number of trials $T_{\max}$ is increased. 

In case a bit flip is required, this node is more involved than the others as two bit estimates need to be flipped simultaneously for the parity constraint to remain satisfied. Two cases can be distinguished. Let $i_{\text{flip}}$ denote the location of the initial bit to be flipped, and $i_{\min_1}$ and $i_{min_2}$ correspond to indices of the least- and second-least-reliable input \glspl{llr}, respectively. If $i_{\text{flip}}=i_{\min_1}$, both $i_{\text{flip}}$ and $i_{min_2}$ get flipped, otherwise, both $i_{\text{flip}}$ and $i_{\min_1}$ get flipped.

\subsection{Error-correction Performance}

\begin{figure}[t]
	\centering
	\input{figures/ec-perf-cmp-flip}\vspace{-6pt}
	\caption{Error-correction performance comparison for a $(512, 128)$ polar code decoded using various 16-bit \gls{crc}-aided \gls{scf}-based algorithms with $T_{\max}=8$ (left) and $T_{\max}=16$ (right). Curves for 16-bit \gls{crc}-aided \gls{scl} decoding with $L \in \{2,4\}$ included for reference.}
	\label{fig:ec-perf-cmp-flip}\vspace{-10pt}
\end{figure}
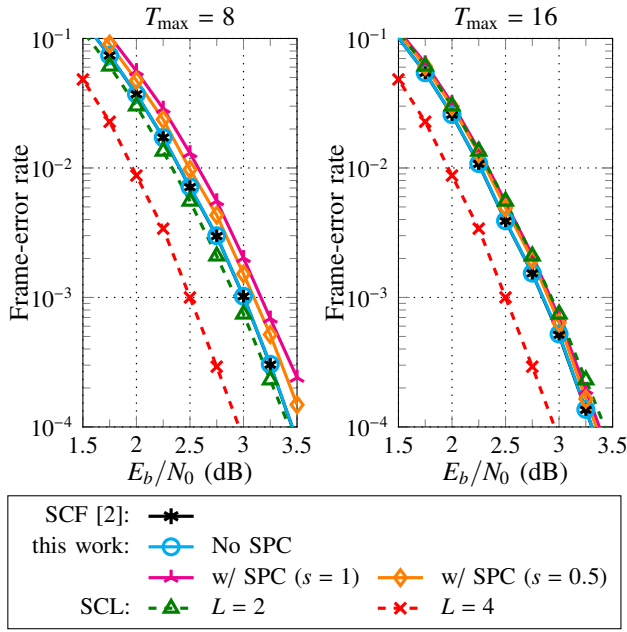

Fig.~\ref{fig:ec-perf-cmp-flip} compares the error-correction performance, in terms of \gls{fer}, for the new decoding algorithm against that of the original \gls{scf} algorithm as proposed by \cite{Afisiadis2014}, where $T_{\max}$ is the maximum number of trials.
A curve for the new algorithm where \gls{spc} nodes are not used is also included for reference. Without \gls{spc} nodes, \gls{spc} codes of length $N_v=4$ (e.g., the node $v$ of Fig.~\ref{fig:fastssc-tree}) are decomposed as a length-2 repetition code combined with a length-2 rate-1 code. Longer \gls{spc} codes generate a chain of rate-$R$ and rate-1 nodes that terminate with a length-2 repetition code combined with a length-2 rate-1 code.
A short $(512,\,128)$ polar code is used with a 16-bit \gls{crc} to be representative of what could be used in next-generation mobile-communication systems \cite{3GPPRANPolar}.
For reference, the figure also shows the \gls{fer} under 16-bit \gls{crc}-aided \gls{scl} decoding for list sizes $L\in\{2,4\}$.
These simulation results are for BPSK-modulated random codewords transmitted over an AWGN channel.

Comparing the \gls{scf} curve (black with asterisk markers) against that of the proposed algorithm without \gls{spc} nodes (cyan with circle markers), either for $T_{\max}=8$ (left) or $16$ (right), it can be seen that the \gls{fer} is virtually the same. 

However, as mentioned in the previous section, the use of (approximate) \gls{spc} nodes incurs a small performance loss. Looking at the results for $T_{\max}=8$ (left side of Fig.~\ref{fig:ec-perf-cmp-flip}) it can be seen that the loss for using \gls{spc} nodes with a scaling factor $s=1$ (purple curve with tri-star markers) amounts slightly more than 0.15\,dB at a \gls{fer} of $10^{-3}$. Setting $s$ to 0.5 (orange curve with diamond markers) reduces that loss to 0.1\,dB at the same \gls{fer}. Increasing $T_{\max}$ to 16 (right side of Fig.~\ref{fig:ec-perf-cmp-flip}) closes the gap between the results with and without the \gls{spc} node: at a \gls{fer} of $10^{-3}$, it is of 0.05\,dB and 0.07\,dB for $s=0.5$ and $s=1$, respectively.

Fig.~\ref{fig:ec-perf-cmp-flip} also shows that, for the simulated (512,\,128) polar code, the error-correction performance under \gls{scf} decoding is in the vicinity of \gls{scl} decoding with a list size $L=2$ (dashed green curve with triangle markers). We note that recent work on \gls{scf} decoding \cite{Condo_WCNC_2018} proposes a low-complexity method, orthogonal to this work, where the error-correction performance for low-rate polar codes of length $N=1024$ was shown to improve by approximately 0.25\,dB.

\section{Hardware Implementation Considerations}\label{sec:hw}
As in \cite{Giard_JETCAS_2017}, the list of decision \glspl{llr} $\mvec{\lambda}$ can be kept sorted with an insert-sort unit, running in parallel with the decoding process, capable of handling a maximum of $\min(P-1,T_{\max}-1)$ input \glspl{llr}, where $P$ is the maximum number of \glspl{llr} that specialized decoders can simultaneously access from memory. By keeping the list sorted, its size can be constrained to $T_{\max}-1$. Thus, a memory of $Q_{\lambda}(T_{\max}-1)$ bits is sufficient to store the decision-\gls{llr} list, where $Q_{\lambda}$ is the number of quantization bits used to represent decision \glspl{llr}. Alongside, a list of the corresponding indices requires a memory of $(T_{\max}-1)\left\lceil\log_2k\right\rceil$\,bits.

Starting with the implementation of the \gls{fastssc} algorithm as described in \cite{Sarkis_JSAC_2014}, the ML unit---an unrolled generic \gls{sc} decoder for length-4-only birepetition codes---would be replaced with a second copy of the unit implementing the repetition node in order to implement the Birepetition node. If $P > T_{\max}-1$, the \gls{spc} unit would need an \gls{llr} sorter to only retain the $T_{\max}-1$ smallest decision \glspl{llr}. Some bit-flipping circuitry needs to be added to all units handling information bits. These modifications are expected to have little impact on the critical path as even the most involved modifications (\gls{spc} unit) only appends bit flips, the remainder of the calculations can occur in parallel.

\subsection*{Latency Comparison}
The only reported hardware implementation of an \gls{scf} decoder \cite{Giard_JETCAS_2017} is built upon a slightly improved semi-parallel \gls{sc} decoder with a latency, in \glspl{cc}, defined as
\begin{equation}\label{eqn:latency:sc}
  \mathcal{L}_{\text{SC}} = 2N+\frac{N}{64}\log_2\left(\frac{N}{256}\right) -
  \sum\limits_{i=0}^{\log_2N} \left\lfloor \frac{b}{2^i}\right\rfloor \left\lceil \frac{2^i}{64} \right\rceil,
\end{equation}
where $N$ is the polar-code length, and $b$ is the location of the first information bit.

The latency of the \gls{fastssc} algorithm cannot be expressed in compact closed form as it heavily depends on the frozen-bit locations, and node types and constraints. However, numerical evaluations show that it is roughly an order of magnitude lower than that of \gls{sc} decoding for all relevant code rates.

\begin{figure}[t]
	\centering
	\input{figures/latency-vs-fer.tex}\vspace{-6pt}
	\caption{Average decoding-execution time for a $(512, 128)$ polar code decoded with 16-bit \gls{crc}-aided \gls{scf}-based algorithms. For this work, $s=0.5$.}
	\label{fig:latency-cmp}\vspace{-10pt}
\end{figure}
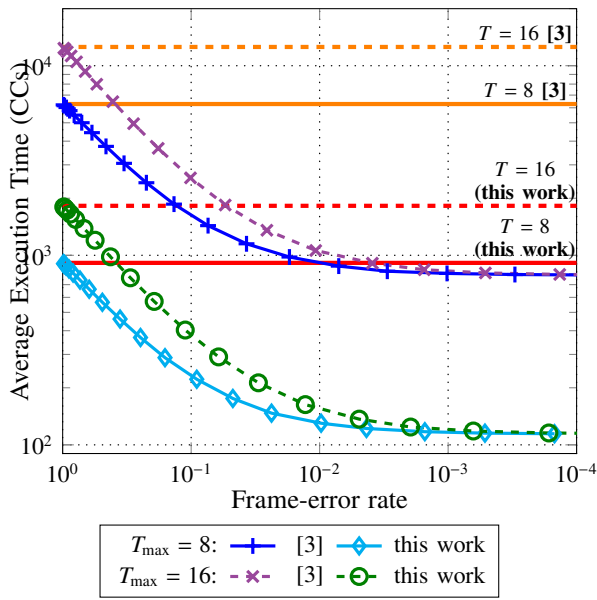

To get a grasp of the improvements under reasonable conditions, Fig.~\ref{fig:latency-cmp} shows a comparison of the average execution time, in terms of clock cycles, between the \gls{scf} decoder of \cite{Giard_JETCAS_2017} with what could be that of the proposed fast-SSC-flip decoding algorithm. For this work, all the original nodes from \cite{Sarkis_JSAC_2014} were used except the ML node which has been replaced with the more efficient Birepetition node. The Repetition, Birepetition, and \gls{spc} nodes are constrained to a maximum size of 32, 64, and 64, respectively. The value of $P$ is set to 64 to match that of \cite{Giard_JETCAS_2017}. All curves are for the same code used in Fig.~\ref{fig:ec-perf-cmp-flip}. The scaling factor $s$ of \eqref{eq:spc:decision} was set to 0.5.

Fig.~\ref{fig:latency-cmp} confirms that the average execution time of the proposed fast-SSC-flip algorithm is close to an order of magnitude lower than that of the other work, both for $T_{\max}=8$ (solid curves) and $T_{\max}=16$ (dashed curves). Furthermore, it can be seen that the worst-case execution time for this work with $T_{\max}=8$ (solid red bottom-most line) is only slightly worse that that of the best-case execution time of \cite{Giard_JETCAS_2017}.

\balance

\section{Conclusion}\label{sec:conclusion}
In this paper, we showed how to merge the state-of-the-art high-speed \gls{sc}-based decoding algorithm---\gls{fastssc}---with the \gls{scf} algorithm. The resulting algorithm was shown to have a significantly higher speed than the \gls{scf}-decoder implementation from the literature while retaining an error-correction performance very close to the original \gls{scf} algorithm.
The key ingredients are the new decision-\gls{llr} calculations and bit-flipping procedures introduced to \gls{scf} decoding for the multi-bit dedicated decoders used in the \gls{fastssc} algorithm.

The proposed decoding algorithm is promising for applications that can handle a variable execution time.
Its error-correction performance can match that of list-based decoding with a small list and its speed tends to that of the fastest low-complexity \gls{sc} decoders in practical operating conditions.

\section*{Acknowledgement}
The authors would like to thank Orion Afisiadis of EPFL for the helpful discussions. This work has been supported by the Swiss National Science Foundation under grant \href{http://p3.snf.ch/project-175813}{\#175813}.

\bibliographystyle{IEEEtran}
\bibliography{IEEEabrv,ConfAbrv,refs}

\end{document}

%% file: figures/pc8.tex
\newcommand{\ubit}[1]{$u_{#1}$}
\newcommand{\fbit}[1]{\color{gray}$u_{#1}$}
\newcommand{\ucw}[1]{$x_{#1}$}
\newcommand{\fcw}[1]{\color{gray}$x_{#1}$}
\newcommand{\ub}[1]{$#1$}
\newcommand{\fb}[1]{\color{gray}$#1$}

\begin{tikzpicture}

\usetikzlibrary{shapes,positioning,arrows,decorations.markings,fit}

\definecolor{varnode_fill}{RGB}{0,0,0}
\definecolor{chknode_fill}{RGB}{255,255,255}

\tikzset{
  chknode/.style={draw,fill=chknode_fill,circle,minimum size=0.3cm, inner sep=0},
  varnode/.style={draw,fill=varnode_fill,circle,minimum size=0.1cm, inner sep=0},
  channel/.style={draw,fill=white,rectangle},
  sep/.style={rectangle,minimum width=0.25cm, inner sep=0},
  empty/.style={rectangle, inner sep=0},
  bit/.style={circle, inner sep = 0}
}

\tikzset{green dotted/.style={draw=green!50!black, line width=1pt,
    dash pattern=on 3pt off 3pt,
    inner sep=0.4mm, rectangle, rounded corners}};

\matrix[row sep=1mm, column sep=1mm] {
  \node[bit] (n0s0) {\fb{u_0}}; & \node[chknode] (n0s1) {$+$}; & \node[sep] (s10) {}; & \node[chknode] (n0s2) {$+$}; & \node[empty] {};              & \node[sep] (s20) {}; & \node[chknode] (n0s3) {$+$}; & \node[empty] {}; & \node[empty] {}; & \node[empty] {}; && \node[bit] (xn0s4) {\ub{x_0}};\\
  \node[bit] (n1s0) {\fb{u_1}}; & \node[varnode] (n1s1) {};    & \node[sep] (s11) {}; &                              & \node[chknode] (n1s2) {$+$};  & \node[sep] (s21) {}; & \node[empty] {};             & \node[chknode] (n1s3) {$+$}; & \node[empty] {}; & \node[empty] {}; && \node[bit] (xn1s4) {\ub{x_1}};\\
  \node[bit] (n2s0) {\ub{u_2}}; & \node[chknode] (n2s1) {$+$}; & \node[sep] (s12) {}; & \node[varnode] (n2s2) {};    & \node[empty] {};              & \node[sep] (s22) {}; & \node[empty] {};             & \node[empty] {}; & \node[chknode] (n2s3) {$+$}; & \node[empty] {}; && \node[bit] (xn2s4) {\ub{x_2}};\\

  \node[bit] (n3s0) {\ub{u_3}}; & \node[varnode] (n3s1) {};    & \node[sep] (s13) {}; & \node[empty] {};             & \node[varnode] (n3s2) {};     & \node[sep] (s23) {}; & \node[empty] {};             & \node[empty] {}; & \node[empty] {}; & \node[chknode] (n3s3) {$+$}; && \node[bit] (xn3s4) {\ub{x_3}};\\

  \node[bit] (n4s0) {\fb{u_4}}; & \node[chknode] (n4s1) {$+$}; & \node[sep] (s14) {}; & \node[chknode] (n4s2) {$+$}; & \node[empty] {};              & \node[sep] (s24) {}; & \node[varnode] (n4s3) {};    & \node[empty] {}; & \node[empty] {}; & \node[empty] {}; && \node[bit] (xn4s4) {\ub{x_4}};\\
  \node[bit] (n5s0) {\ub{u_5}}; & \node[varnode] (n5s1) {};    & \node[sep] (s15) {}; &                              & \node[chknode] (n5s2) {$+$};  & \node[sep] (s25) {}; & \node[empty] {};             & \node[varnode] (n5s3) {}; & \node[empty] {}; &  \node[empty] {}; && \node[bit] (xn5s4) {\ub{x_5}};\\
  \node[bit] (n6s0) {\ub{u_6}}; & \node[chknode] (n6s1) {$+$}; & \node[sep] (s16) {}; & \node[varnode] (n6s2) {};    & \node[empty] {};              & \node[sep] (s26) {}; & \node[empty] {};             & \node[empty] {}; & \node[varnode] (n6s3) {}; &  \node[empty] {}; && \node[bit] (xn6s4) {\ub{x_6}};\\
  
  \node[bit] (n7s0) {\ub{u_7}}; & \node[varnode] (n7s1) {};    & \node[sep] (s17) {}; &                              & \node[varnode] (n7s2) {};  & \node[sep] (s27) {}; & \node[empty] {};             & \node[empty] {}; & \node[empty] {}; &  \node[varnode] (n7s3) {}; && \node[bit] (xn7s4) {\ub{x_7}};\\
};
\path[-] (n0s0) edge (n0s1) (n0s1) edge (n0s2) (n0s2) edge (n0s3) (n0s3) edge (xn0s4);
\path[-] (n1s0) edge (n1s1) (n1s1) edge (n1s2) (n1s2) edge (n1s3) (n1s3) edge (xn1s4);
\path[-] (n2s0) edge (n2s1) (n2s1) edge (n2s2) (n2s2) edge (n2s3) (n2s3) edge (xn2s4);
\path[-] (n3s0) edge (n3s1) (n3s1) edge (n3s2) (n3s2) edge (n3s3) (n3s3) edge (xn3s4);
\path[-] (n4s0) edge (n4s1) (n4s1) edge (n4s2) (n4s2) edge (n4s3) (n4s3) edge (xn4s4);
\path[-] (n5s0) edge (n5s1) (n5s1) edge (n5s2) (n5s2) edge (n5s3) (n5s3) edge (xn5s4);
\path[-] (n6s0) edge (n6s1) (n6s1) edge (n6s2) (n6s2) edge (n6s3) (n6s3) edge (xn6s4);
\path[-] (n7s0) edge (n7s1) (n7s1) edge (n7s2) (n7s2) edge (n7s3) (n7s3) edge (xn7s4);

\path[-] (n0s1) edge (n1s1);
\path[-] (n2s1) edge (n3s1);
\path[-] (n4s1) edge (n5s1);
\path[-] (n6s1) edge (n7s1);

\path[-] (n0s2) edge (n2s2);
\path[-] (n1s2) edge (n3s2);
\path[-] (n4s2) edge (n6s2);
\path[-] (n5s2) edge (n7s2);

\path[-] (n0s3) edge (n4s3);
\path[-] (n1s3) edge (n5s3);
\path[-] (n2s3) edge (n6s3);
\path[-] (n3s3) edge (n7s3);

\node (g_n1s2) [green dotted, fit = (n4s2) (n5s2) (n6s2) (n7s2)] {};

\end{tikzpicture}

%% file: figures/sc-tree.tex
\begin{tikzpicture}[baseline = (0_7.center),
        level/.style={level distance = 6mm},
        level 1/.style={sibling distance=19mm, edge from parent/.style={draw,blue,line width=1.5pt}},
        level 2/.style={sibling distance=9.5mm, edge from parent/.style={draw,blue,line width=1pt}},
        level 3/.style={sibling distance=4.7mm, edge from parent/.style={draw,blue,line width=0.5pt}},
        ]

\tikzset{
frozen/.style={thick,draw=black,fill=white,minimum size=3mm,circle, inner sep=0},
fullspace/.style={thick,draw=black,fill=black,minimum size=3mm,circle, inner sep = 0},
mixed/.style={thick,draw=black,fill=gray,minimum size=3mm,circle, inner sep = 0},
phantom/.style={draw=white,fill=white,minimum size=3mm,circle, inner sep = 0},
}

\tikzset{
parallel segment/.style={
   segment distance/.store in=\segDistance,
   segment pos/.store in=\segPos,
   segment length/.store in=\segLength,
   to path={
   ($(\tikztostart)!\segPos!(\tikztotarget)!\segLength/2!(\tikztostart)!\segDistance!90:(\tikztotarget)$) -- 
   ($(\tikztostart)!\segPos!(\tikztotarget)!\segLength/2!(\tikztotarget)!\segDistance!-90:(\tikztostart)$)  \tikztonodes
   }, 
   segment pos=.5,
   segment length=2.5ex,
   segment distance=-1mm,
},
}

\tikzset{green dotted/.style={draw=green!50!black, line width=0.75pt,
    dash pattern=on 3pt off 3pt,
    inner sep=0.4mm, rectangle, rounded corners}};

\node[mixed] (p){} [grow=left]
	child {node[mixed] (2_0){}
		child {node[mixed] (1_0){}
			child {node[frozen] (a0_0){}
			}
			child {node[frozen] (a0_1){} edge from parent[red]
			}
		}
		child {node[mixed] (1_2){} edge from parent[red]
			child {node[fullspace] (0_2){}
			}
			child {node[fullspace] (0_3){} edge from parent[red]
			}
		}
	}
	child {node[mixed] (v){\rotatebox{-90}{\textcolor{white}{$v$}}} edge from parent[red]
		child {node[mixed] (cl){}
			child {node[frozen] (0_4){}
			}
			child {node[fullspace] (0_5){} edge from parent[red]
			}
		}
		child {node[mixed] (cr){} edge from parent[red]
			child {node[fullspace] (0_6){}
			}
			child {node[fullspace] (0_7){} edge from parent[red]
			}
		}
	}
;

\draw[->,line width=0.65pt] (p) to[parallel segment,segment length=4ex] node[above left=-2.0mm] {\rotatebox{-90}{\footnotesize $\alpha_v$}} (v);
\draw[->,line width=0.65pt] (v) to[parallel segment,segment length=4ex] node[below right=-2.0mm] {\rotatebox{-90}{\footnotesize $\beta_v$}} (p);

\draw[->,line width=0.65pt] (cl) to[parallel segment] (v) {};
\node at ($(cl.south)-(0,0.18)$) {\rotatebox{-90}{\footnotesize $\beta_l$}};
\draw[->,line width=0.65pt] (v) to[parallel segment] node[above right=-2.0mm] {\rotatebox{-90}{\footnotesize $\alpha_l$}} (cl);

\draw[->,line width=0.65pt] (v) to[parallel segment] (cr) {};
\node at ($(cr.north)+(-0.04,0.16)$) {\rotatebox{-90}{\footnotesize $\alpha_r$}};
\draw[->,line width=0.65pt] (cr) to[parallel segment] node[below right=-2.0mm] {\rotatebox{-90}{\footnotesize $\beta_r$}} (v);

\node (g_concat) [green dotted, fit = (v)] {};

\end{tikzpicture}

%% file: figures/fastssc-tree.tex
\begin{tikzpicture}[baseline = (v.center),
        level/.style={level distance = 6mm},
        level 1/.style={sibling distance=19mm, edge from parent/.style={draw,blue,line width=1.5pt}},
        level 2/.style={sibling distance=9.5mm, edge from parent/.style={draw,blue,line width=1pt}},
        level 3/.style={sibling distance=4.7mm, edge from parent/.style={draw,blue,line width=0.5pt}},
        ]

\tikzset{
frozen/.style={thick,draw=black,fill=white,minimum size=3mm,circle, inner sep=0},
fullspace/.style={thick,draw=black,fill=black,minimum size=3mm,circle, inner sep = 0},
mixed/.style={thick,draw=black,fill=gray,minimum size=3mm,circle, inner sep = 0},
phantom/.style={draw=white,fill=white,minimum size=3mm,circle, inner sep = 0},
birep/.style={thick,draw=black,pattern=north east lines,pattern color=purple,minimum size=3mm,circle, inner sep = 0},
spc/.style={thick,draw=black,pattern=crosshatch,pattern color=orange,minimum size=3mm,circle, inner sep = 0},
}

\tikzset{green dotted/.style={draw=green!50!black, line width=1pt,
    dash pattern=on 3pt off 3pt,
    inner sep=0.4mm, rectangle, rounded corners}};

\node[mixed] (p){} [grow=left]
	child {node[birep] (2_0){}
	}
	child {node[spc, label=right:\rotatebox{-90}{$v$}] (v){} edge from parent[red]
	}
;

\node (g_concat) [green dotted, fit = (v)] {};

\end{tikzpicture}

%% file: figures/ec-perf-cmp-flip.tex
\usetikzlibrary{plotmarks}

\definecolor{darkgreen}{RGB}{0, 128, 0}
\definecolor{mypurple}{RGB}{153, 71, 155}

\begin{tikzpicture}

  \pgfplotsset{
    grid style = {
      dash pattern = on 0.05mm off 1mm,
      line cap = round,
      black,
      line width = 0.5pt
    },
    title style = {font=\fontsize{10pt}{7.2}\selectfont},
    label style = {font=\fontsize{10pt}{7.2}\selectfont},
    tick label style = {font=\fontsize{9pt}{7.2}\selectfont}
  }

  \begin{semilogyaxis}[%
    title={$T_{\max}=8$}, title style={yshift=-0.6em},
    xlabel=$E_b/N_0$ (dB),xtick={1,1.5,2,...,4},%
    xlabel style={yshift=0.4em},%
    minor x tick num={1},
    xmin=1.5,xmax=3.5,%
    ymin=1e-4,ymax=1e-1,%
    ylabel=Frame-error rate, ylabel style={yshift=-0.6em},%
    width=0.5\columnwidth, height=6.75cm, grid=major,%
    legend style={
      anchor={center},
      cells={anchor=west},
      column sep=1.5mm,
      font=\fontsize{9pt}{7.2}\selectfont,
      mark size=3.0pt,
      mark options=solid
    },
    legend columns=3,
    legend to name=ec-perf-cmp-flip-legend,
    mark size=3.0pt,
    mark options=solid]
    
    \addlegendimage{empty legend}
    \addlegendentry[anchor=east]{SCF \cite{Afisiadis2014}:}
    \addplot[very thick,color=black,mark=asterisk] table[x=EbN0dB,y=FER] {data/512.128.s1.029.float.scf.t8.c16.csv};
    \addlegendentry{}
    \addlegendimage{empty legend}
    \addlegendentry{}

    \addlegendimage{empty legend}
    \addlegendentry[anchor=east]{this work:}

    \addplot[very thick,color=cyan, mark=o] table[x=EbN0dB,y=FER] {data/512.128.s1.029.float.fastsscf.t8.c16.ssc+rep+r1+drep.csv};
    \addlegendentry{No SPC}
    \addlegendimage{empty legend}
    \addlegendentry{}

    \addlegendimage{empty legend}
    \addlegendentry[anchor=east]{\phantom{this work:}}

    \addplot[very thick,color=magenta, mark=Mercedes star] table[x=EbN0dB,y=FER] {data/512.128.s1.029.float.fastsscf.t8.c16.ssc+rep+r1+rspc+0spc+drep+repspc.csv};
    \addlegendentry{w/ SPC ($s=1$)}

    \addplot[very thick,color=orange, mark=diamond] table[x=EbN0dB,y=FER] {data/512.128.s1.029.float.fastsscf.t8.c16.ssc+rep+r1+rspc+0spc+drep+repspc.0.5.csv};
    \addlegendentry{w/ SPC ($s=0.5$)}

    \addlegendimage{empty legend}
    \addlegendentry[anchor=east]{SCL:}

    \addplot[very thick,dashed,color=darkgreen, mark=triangle] table[x=EbN0dB,y=FER] {data/512.128.s1.029.float.scl.l2.c16.csv};
    \addlegendentry{$L = 2$}

    \addplot[very thick,dashed,color=red, mark=x] table[x=EbN0dB,y=FER] {data/512.128.s1.029.float.scl.l4.c16.csv};
    \addlegendentry{$L = 4$}

  \end{semilogyaxis}
\end{tikzpicture}\hspace{-5pt}
\begin{tikzpicture}

  \pgfplotsset{
    grid style = {
      dash pattern = on 0.05mm off 1mm,
      line cap = round,
      black,
      line width = 0.5pt
    },
    title style = {font=\fontsize{10pt}{7.2}\selectfont},
    label style = {font=\fontsize{10pt}{7.2}\selectfont},
    tick label style = {font=\fontsize{9pt}{7.2}\selectfont}
  }

  \begin{semilogyaxis}[%
    title={$T_{\max}=16$}, title style={yshift=-0.6em},
    xlabel=$E_b/N_0$ (dB),xtick={1,1.5,2,...,4},%
    xlabel style={yshift=0.4em},%
    minor x tick num={1},
    xmin=1.5,xmax=3.5,%
    ymin=1e-4,ymax=1e-1,%
    ylabel=Frame-error rate, ylabel style={yshift=-0.6em},%
    width=0.5\columnwidth, height=6.75cm, grid=major,%
    mark size=3.0pt,
    mark options=solid]
    
    \addplot[very thick,color=black,mark=asterisk] table[x=EbN0dB,y=FER] {data/512.128.s1.029.float.scf.t16.c16.csv};

    \addplot[very thick,color=cyan, mark=o] table[x=EbN0dB,y=FER] {data/512.128.s1.029.float.fastsscf.t16.c16.ssc+rep+r1+drep.csv};

    \addplot[very thick,color=magenta, mark=Mercedes star] table[x=EbN0dB,y=FER] {data/512.128.s1.029.float.fastsscf.t16.c16.ssc+rep+r1+rspc+0spc+drep+repspc.csv};

    \addplot[very thick,color=orange, mark=diamond] table[x=EbN0dB,y=FER] {data/512.128.s1.029.float.fastsscf.t16.c16.ssc+rep+r1+rspc+0spc+drep+repspc.0.5.csv};

    \addplot[very thick,dashed,color=darkgreen, mark=triangle] table[x=EbN0dB,y=FER] {data/512.128.s1.029.float.scl.l2.c16.csv};

    \addplot[very thick,dashed,color=red, mark=x] table[x=EbN0dB,y=FER] {data/512.128.s1.029.float.scl.l4.c16.csv};

  \end{semilogyaxis}
\end{tikzpicture}
\\
\ref{ec-perf-cmp-flip-legend}

%% file: figures/latency-vs-fer.tex
\usetikzlibrary{plotmarks}

\definecolor{darkgreen}{RGB}{0, 128, 0}
\definecolor{mypurple}{RGB}{153, 71, 155}

\begin{tikzpicture}

  \pgfplotsset{
    grid style = {
      dash pattern = on 0.05mm off 1mm,
      line cap = round,
      black,
      line width = 0.5pt
    },
    label style = {font=\fontsize{10pt}{7.2}\selectfont},
    tick label style = {font=\fontsize{9pt}{7.2}\selectfont}
  }

  \tikzset{
    branch/.style={fill,shape=circle,scale=0.4},
  }

 \begin{loglogaxis}[%
   xlabel={Frame-error rate},xtick=data,%
   xlabel style={yshift=0.4em},%
   log basis x=10,%
   x dir=reverse, xmax=1e-0, xmin=1e-4,%
   xtick={1e0,1e-1,1e-2,1e-3,1e-4},
   ymin=9e1,ymax=2e4,
   ylabel={Average Execution Time (CCs)}, ylabel style={yshift=-1.2em},%
   width=0.95\columnwidth, height=7.5cm, grid=major,%
   legend style={
     anchor={center},
     cells={anchor=west},
     column sep=0.75mm,
     font=\fontsize{9pt}{7.2}\selectfont,
     mark size=3.0pt,
     mark options=solid
   },
   legend columns=3,
   legend to name=latency-legend,
   mark size=3.0pt,
   mark options=solid
   ]
    
   \addlegendimage{empty legend}
   \addlegendentry[anchor=east]{$T_{\max}=8$:}

   \draw [ultra thick, draw=orange] (axis cs: 1e0,6280) -- (axis cs: 1e-4,6280) node[pos=0.9, above, yshift=-1mm] {\footnotesize\bf $T=8$ \cite{Giard_JETCAS_2017}};
   \draw [ultra thick, dashed, draw=orange] (axis cs: 1e0,12560) -- (axis cs: 1e-4,12560) node[pos=0.9, above, yshift=-1mm] {\footnotesize\bf $T=16$ \cite{Giard_JETCAS_2017}};

   \draw [ultra thick, draw=red] (axis cs: 1e0,912) -- (axis cs: 1e-4,912) node[pos=0.9, above, yshift=-1mm] {\footnotesize\bf \shortstack{$T=8$\\(this work)}};
   \draw [ultra thick, dashed, draw=red] (axis cs: 1e0,1824) -- (axis cs: 1e-4,1824) node[pos=0.9, above, yshift=-1mm] {\footnotesize\bf \shortstack{$T=16$\\(this work)}};
        
   \addplot[very thick,blue,mark=+] table[x=FER,y expr=\thisrow{avgTrials}*785] {data/512.128.s1.029.float.scf.t8.c16.csv};
    \addlegendentry{\cite{Giard_JETCAS_2017}}

    \addplot[very thick,color=cyan, mark=diamond] table[x=FER,y expr=\thisrow{avgTrials}*114] {data/512.128.s1.029.float.fastsscf.t8.c16.ssc+rep+r1+rspc+0spc+drep+repspc.0.5.csv};
    \addlegendentry{this work}

   \addlegendimage{empty legend}
   \addlegendentry[anchor=east]{$T_{\max}=16$:}
   \addplot[very thick,dashed,mypurple, mark=x] table[x=FER,y expr=\thisrow{avgTrials}*785] {data/512.128.s1.029.float.scf.t16.c16.csv};
   \addlegendentry{\cite{Giard_JETCAS_2017}}

   \addplot[very thick,dashed,color=darkgreen, mark=o] table[x=FER,y expr=\thisrow{avgTrials}*114] {data/512.128.s1.029.float.fastsscf.t16.c16.ssc+rep+r1+rspc+0spc+drep+repspc.0.5.csv};
   \addlegendentry{this work}

 \end{loglogaxis}

\end{tikzpicture}
\\\vspace{3pt}
\ref{latency-legend}